\documentclass[aps,prd,preprint,preprintnumbers,nofootinbib,superscriptaddress]{revtex4}
\usepackage{ae}
\usepackage{bm} 
\usepackage{color}
\usepackage{amssymb}
\usepackage{amsmath}
\usepackage{amsfonts}
\usepackage{graphicx}
\usepackage{slashed}
\usepackage{wasysym}
\usepackage{setspace}
\usepackage{ulem}
\usepackage{multirow}

\newcommand{\Eqref}[1]{Eq.~\eqref{#1}}

\newcommand{\rk}{{\rm k}}
\newcommand{\rx}{{\rm x}}
\newcommand{\ry}{{\rm y}}
\newcommand{\rz}{{\rm z}}
\newcommand{\ri}{{\rm i}}

\renewcommand{\vec}[1]{\mathbf{#1}}
\renewcommand{\frak}[1]{\mathfrak{#1}}

\allowdisplaybreaks

\begin{document}

\setlength{\unitlength}{1mm}

\title{Probing vacuum birefringence using x-ray free electron and optical high-intensity lasers}

\author{Felix Karbstein}\email{felix.karbstein@uni-jena.de}
\affiliation{Helmholtz-Institut Jena, Fr\"obelstieg 3, 07743 Jena, Germany}
\affiliation{Theoretisch-Physikalisches Institut, Abbe Center of Photonics, \\ Friedrich-Schiller-Universit\"at Jena, Max-Wien-Platz 1, 07743 Jena, Germany}
\author{Chantal Sundqvist}\email{chantal.sundqvist@uni-jena.de}
\affiliation{Helmholtz-Institut Jena, Fr\"obelstieg 3, 07743 Jena, Germany}
\affiliation{Theoretisch-Physikalisches Institut, Abbe Center of Photonics, \\ Friedrich-Schiller-Universit\"at Jena, Max-Wien-Platz 1, 07743 Jena, Germany}

\date{\today}

\begin{abstract}
 Vacuum birefringence is one of the most striking predictions of strong field quantum electrodynamics:
Probe photons traversing a strong field region can indirectly sense the applied ``pump'' electromagnetic field via quantum fluctuations of virtual charged particles which couple to both pump and probe fields.
This coupling is sensitive to the field alignment and can effectively result in two different indices of refraction for the probe photon polarization modes giving rise to a birefringence phenomenon. 
In this article we perform a dedicated theoretical analysis of the proposed discovery experiment of vacuum birefringence at a x-ray free electron laser/optical high-intensity laser facility.
Describing both pump and probe laser pulses realistically in terms of their macroscopic electromagnetic fields, we go beyond previous analyses by accounting for various effects not considered before in this context.
Our study facilitates stringent quantitative predictions and optimizations of the signal in an actual experiment.
\end{abstract}

\maketitle

\section{Introduction}

The quantum vacuum amounts to a highly nontrivial state, characterized by the omnipresence of fluctuations of virtual particles.
While the microscopic theory of quantum electrodynamics (QED) does not provide for a direct (tree-level) interaction among photons, effective interactions of this kind are induced by quantum fluctuations of charged particles -- in QED, electrons and positrons.
Specifically in strong electromagnetic fields quantum fluctuations give rise to effective, nonlinear interactions among photons and macroscopic electromagnetic fields \cite{Euler:1935zz,Heisenberg:1935qt,Weisskopf} (for reviews, see \cite{Dittrich:1985yb,Dittrich:2000zu,Marklund:2008gj,Dunne:2008kc,Heinzl:2008an,DiPiazza:2011tq,Dunne:2012vv,Battesti:2012hf,King:2015tba}).
One of the most famous optical signatures of QED vacuum nonlinearity is vacuum birefringence \cite{Toll:1952,Baier,BialynickaBirula:1970vy,Adler:1971wn,Kotkin:1996nf}
experienced by probe photons traversing a strong field region, which is actively searched for in precision experiments using macroscopic magnetic fields \cite{Cantatore:2008zz,Berceau:2011zz}; see \cite{Zavattini:2016sqz} for a recent proposal.

At zero field, the vacuum is characterized by translational invariance and the absence of any preferred direction. 
Conversely, an external electromagnetic field generically introduces a preferred direction, and, in the presence of inhomogeneities, also breaks translational invariance for charged particles.
Via the charged particle-antiparticle fluctuations coupling to the external electromagnetic field, this preferred direction can also impact probe photon propagation.
It can in particular affect the two probe photon polarization modes differently, and thereby effectively result in two different indices of refraction for these polarization modes. This can give rise to a birefringence phenomenon, manifesting itself in a tiny ellipticity picked up by an originally purely linearly polarized probe photon beam.
As an ellipticity signal has a nonvanishing overlap with both linearly independent polarization modes spanning the transverse probe photon polarizations,
vacuum birefringence alternatively manifests itself in signal photons scattered into an -- originally empty -- perpendicularly polarized mode.
The number of perpendicularly polarized photons constitutes the most straightforward signature to be observed experimentally in a high-intensity laser experiment aiming at an experimental verification of vacuum birefringence put forward by \cite{Heinzl:2006xc}, envisioning the combination of an optical high-intensity laser as pump and a linearly polarized x-ray pulse as probe; cf. also \cite{DiPiazza:2006pr,Dinu:2013gaa}.
For proposals of vacuum birefringence experiments with dipole, synchrotron and gamma radiation, cf. \cite{Kotkin:1996nf,Ilderton:2016khs,King:2016jnl}.

Other theoretical proposals aiming at the experimental study of optical signatures of quantum vacuum nonlinearity
have focused on interference effects \cite{King:2013am,Tommasini:2010fb,Hatsagortsyan:2011}, photon-photon
scattering in the form of laser-pulse collisions
\cite{Lundstrom:2005za,Lundin:2006wu,King:2012aw}, quantum reflection \cite{Gies:2013yxa}, as well as photon merging \cite{Yakovlev:1966,DiPiazza:2007cu,Gies:2014jia,Gies:2016czm}
and splitting 
\cite{BialynickaBirula:1970vy,Adler:1971wn,Adler:1970gg,Papanyan:1971cv,Stoneham:1979,Baier:1986cv,Adler:1996cja,DiPiazza:2007yx}.

In a recent feasibility study for detecting QED vacuum birefringence with x-ray free electron lasers (FELs) and high-power optical lasers, the probe photons are assumed to traverse the pump field on straight lines resembling their trajectories in vacuum at zero field \cite{Schlenvoigt:2016}.
The ellipticity acquired by a probe photon counter-propagating the pump at a given impact parameter is then traced on such a straight-line trajectory. 
By construction this procedure only gives rise to signal photons emitted exactly in forward direction.
The conclusion of this study was that the experimental verification of vacuum birefringence with state-of-the-art laser systems and polarimetry is possible, but
-- as the measurement is dominated by noise -- requires a large number of laser shots to increase the statistics. 

However, another recent publication \cite{Karbstein:2015xra} emphasized a way to overcome the noise domination, namely by exploiting the scattering of signal photons out of the cone of the incident probe beam.
Modelling the probe pulse as a non-divergent macroscopic electromagnetic field of finite transverse extent, the signal photons induced upon transversal of the pump laser field genuinely feature a finite divergence.
Assuming the probe in the interaction region to amount to the essentially divergence free segment of a Gaussian beam around its focus, by an adequate choice of the beam parameters the divergence of the signal photons can be made substantially larger than the far-field divergence of the original probe beam.
Looking for signal photons scattered outside the divergence of the probe beam, the demands on the polarization purity are less stringent for detection under such angles due to the significantly lower background photon flux.
In \cite{Karbstein:2015xra} this scenario was studied under idealized conditions, assuming the beam axes of the pump and the probe to be perfectly aligned and the pulses to be exactly counter-propagating.
While the pump was realistically modeled as a Gaussian laser pulse in the paraxial approximation, the description of the probe was less elaborate and only its essential features were explicitly accounted for.
More specifically, it was modeled as a plane wave with a longitudinal envelope implementing a finite pulse duration.
Its finite transverse extent was only indirectly accounted for, which limited the discussion to certain special cases, namely probe beams either significantly narrower or wider than the pump beam.

Here we go beyond these limitations and consider probe beams of finite width and generic elliptically shaped cross-sections. Besides, we account for several additional parameters of experimental relevance, like a finite angle between the beams' axes and finite impact parameters.
These improvements facilitate unprecedented theoretical predictions of the experimental signals attainable in a dedicated discovery experiment of QED vacuum birefringence at a
FEL/high-intensity laser facility, like the upcoming Helmholtz International Beamline for Extreme Fields (HIBEF) \cite{HIBEF} at the European XFEL \cite{XFEL} at DESY.

Our article is structured as follows:
After recalling the interpretation of vacuum birefringence in terms of a vacuum emission process in Sec.~\ref{sec:VacEm},
in Sec.~\ref{sec:specconf} we detail the specific pump and probe field configuration considered in this article.
We aim at a realistic description of the pump and probe laser pulses available in the laboratory. 
To this end we account for various experimentally relevant effects, such as, e.g., finite impact parameters and collision angles.
Section~\ref{sec:results} is devoted to our results for the differential number of signal photons.
After a thorough discussion of the generic case, we specialize to a counter-propagation geometry of the pump and probe laser pulses.
The latter geometry is of particular relevance as it allows for the maximum number of signal photons.
In Sec.~\ref{sec:predictions} we provide explicit predictions for the numbers of perpendicularly polarized signal photons which could be measured with state-of-the-art technology. To this end, we numerically evaluate our result for the case of exact counter propagation and vanishing offset parameters for various probe beam cross-sections.
As a real experiment always suffers from shot-to-shot variations, manifesting itself, e.g., in a nonzero impact parameter, the explicit values for the numbers of perpendicularly polarized signal photons obtained here can be considered as the prospective numbers of signal photons attainable under optimal experimental conditions. Our general formula is capable to also account for these shot-to-shot variations, as needed for a concrete design study.
Finally, we end with conclusions and a outlook in Sec.~\ref{sec:Con&Out}.

\section{Vacuum birefringence as vacuum emission process}\label{sec:VacEm}

Following \cite{Karbstein:2014fva,Karbstein:2015qwa} we formally consider the signal photons induced in the interaction volume as emitted from the quantum vacuum subjected to the macroscopic electromagnetic fields of the pump and probe laser pulses.
In this framework, the signal of vacuum birefringence amounts to signal photons polarized perpendicularly to the incident probe photon beam.
The initial state containing no signal photons is denoted by $|0\rangle$. In the presence of inhomogeneous electromagnetic fields, the quantum fluctuations of charged particles can trigger transitions to states with signal photons.
Aiming at the study of vacuum birefringence in an FEL/high-intensity laser scenario it suffices to consider transitions to single signal photon states $|\gamma_{p'}(\vec{k}')\rangle\equiv a^\dag_{\vec{k}',p'}|0\rangle$ only. Here $p'\in\{1,2\}$ labels the polarization of the induced signal photon of four wave-vector $k'^\mu=\rk'\hat{k}^\mu$, where $\rk'=|\vec{k}'|$, $\hat k^\mu=(1,\hat{\vec{k}}')$ and $\hat{\vec{k}}'=\vec{k}'/\rk'$ is the unit wave-vector.
Transitions to multi-photon states are strongly suppressed.

The zero-to-single signal photon transition amplitude stimulated by a macroscopic, slowly varying electromagnetic field $F^{\mu\nu}(x)$ is given by \cite{Karbstein:2014fva}
\begin{equation}
 {\cal S}_{(p')}(\vec{k}')\equiv\langle\gamma_{p'}(\vec{k}')|\int{\rm d}^4x\, f^{\mu\nu}(x)\frac{\partial{\cal L}}{\partial F^{\mu\nu}}(x)|0\rangle\,, \label{eq:Sp1}
\end{equation}
where $f^{\mu\nu}(x)$ is the field strength tensor of the second-quantized signal photon field,
and $\cal L$ is the Heisenberg-Euler effective Lagrangian \cite{Heisenberg:1935qt}, encoding quantum corrections to Maxwell's theory of classical electrodynamics.
The differential number of signal photons with polarization $p'$ to be detected far outside the interaction volume is then determined with Fermi's golden rule, 
\begin{equation}
 {\rm d}^3N_{(p')}=\frac{{\rm d}^3k'}{(2\pi)^3}\bigl|{\cal S}_{(p')}(\vec{k}')\bigr|^2\,. \label{eq:Fermi}
\end{equation}

Let us emphasize that in our context the field strength tensor $F^{\mu\nu}(x)$ accounts for the combined macroscopic electric $\vec{E}(x)$ and magnetic $\vec{B}(x)$ fields of both the pump and probe laser pulses; $F^{0i}=-F^{i0}=E_i$ and $F^{ij}=\epsilon_{ijk}B_k$.
We use the Heaviside-Lorentz System and units where $c=\hbar=1$; our metric convention is $g^{\mu\nu}=\rm{diag}(-,+,+,+)$ and $\alpha=\frac{e^2}{4\pi}\approx\frac{1}{137}$.

\begin{figure}
\center
\includegraphics[width=0.3\textwidth]{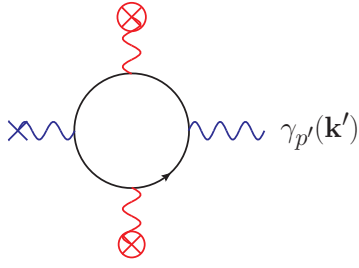}
\caption{Feynman diagram of the leading-order process evaluated in \Eqref{eq:Sp2}, inducing signal photons of wave vector $\vec{k}'$ and polarization $p'$.
The process is cubic in the combined macroscopic electromagnetic fields of the pump and probe pulses represented by wiggly lines ending at crosses. The dominant process features a single coupling to the x-ray probe ($\times$) and a quadratic coupling to the high-intensity pump ($\otimes$) fields.}
\label{fig:Feyndiag}
\end{figure}

If the amplitudes of $\vec{E}(x)$ and $\vec{B}(x)$ are substantially smaller than the {\it critical} electric (magnetic) field strength $E_{\rm cr}=\frac{m^2c^3}{\hbar e}\approx1.3\cdot10^{18}\,{\rm V}/{\rm m}$ ($B_{\rm cr}=\frac{E_{\rm cr}}{c}\approx4.4\cdot10^9\,{\rm T}$),
which is true for all present and near future high-intensity laser systems, \Eqref{eq:Sp1} can be represented in a particular compact form.
The leading contribution to \Eqref{eq:Sp1} in this limit depicted in Fig.~\ref{fig:Feyndiag} is given by
\begin{equation}
{\cal S}_{(p')}(\vec{k}')
 =\frac{{\rm i}\sqrt{\alpha}}{45}\frac{m^2}{4\pi^{\frac{3}{2}}}\frac{1}{\sqrt{2\rk'}} \Bigl(\frac{e}{m^2}\Bigr)^3\int{\rm d}^4x\,{\rm e}^{{\rm i}k'x} \Bigl[4{\cal F}(x)F_{\mu\nu}(x) +7{\cal G}(x){}^*F_{\mu\nu}(x) \Bigr]\hat f^{\mu\nu}_{(p')}(k') \,.
 \label{eq:Sp2}
\end{equation}
Here $\hat f^{\mu\nu}_{(p')}(k')=k'^\mu\epsilon^{\nu}_{(p')}(\hat{\vec{k}}')- k'^\nu\epsilon^{\mu}_{(p')}(\hat{\vec{k}}')$ denotes the normalized signal-photon field strength tensor in momentum space
and ${\cal F}=\frac{1}{4}F_{\mu\nu}F^{\mu\nu}=\frac{1}{2}(\vec{B}^2-\vec{E}^2)$, ${\cal G}=\frac{1}{4}F_{\mu\nu}{}^*F^{\mu\nu}=-\vec{E}\cdot\vec{B}$ are the gauge and Lorentz invariants of the electromagnetic field;
$^*F^{\mu\nu}=\frac{1}{2}\epsilon^{\mu\nu\alpha\beta}F_{\alpha\beta}$ is the dual field strength tensor ($\epsilon^{0123}=1$),
and we employ the short-hand notation $k'x=k'_\mu x^\mu$.

In spherical coordinates we have $\hat{\vec{k}}'=(\cos\varphi'\sin\vartheta',-\sin\varphi'\sin\vartheta',-\cos\vartheta')$.
Note that our conventions are such that for $\vartheta'=0$ we have $\hat{\vec{k}}'|_{\vartheta'=0}=-\hat{\vec{e}}_{\rm z}$.
Besides, it is convenient to introduce the vector
\begin{equation}
\hat{\vec{e}}_{\varphi',\vartheta',\beta'}=
\left(\begin{array}{c}
  \sin\varphi'\sin\beta'-\cos\varphi'\cos\vartheta'\cos\beta' \\
  \cos\varphi'\sin\beta'+\sin\varphi'\cos\vartheta'\cos\beta' \\
  -\sin\vartheta'\cos\beta'
 \end{array}\right), \label{eq:epsilons1}
\end{equation}
which -- by means of $\beta'$ -- parameterizes all unit vectors normal to $\hat{\vec{k}}'$. 
We use it to define $\epsilon^\mu_{(p')}(\hat{\vec{k}}')=(0,\hat{\vec{e}}_{\varphi',\vartheta',\beta'})$,
which can be employed to span the two transverse photon polarization modes fulfilling $\hat{k}'_\mu\epsilon^\mu_{(p')}(\hat{\vec{k}}')=0$.
Sticking to linear polarizations, without loss of generality $\epsilon^\mu_{(1)}(\hat{\vec{k}}')$ is fixed by a particular choice of $\beta'$,
and the perpendicular vector by $\epsilon^\mu_{(2)}(\hat{\vec{k}}')=\epsilon^\mu_{(1)}(\hat{\vec{k}}')|_{\beta'\to\beta'-\frac{\pi}{2}}$.

\begin{figure}
\center
\includegraphics[width=0.74\textwidth]{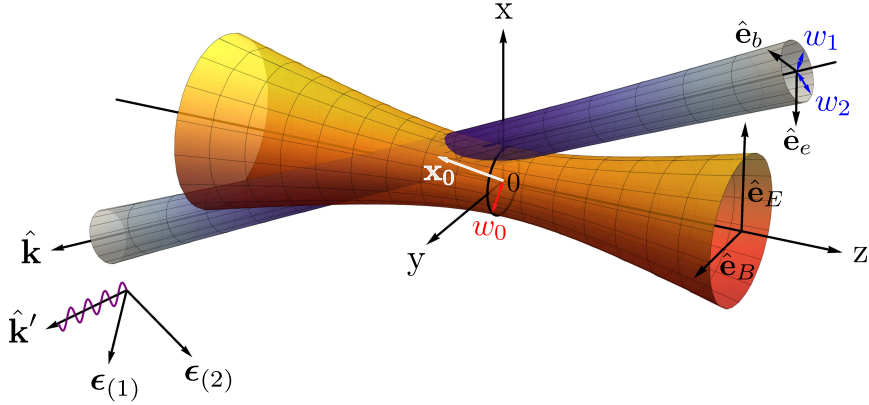}
\vspace*{-0.3cm}
\caption{Sketch of the scenario considered here. 
The transverse profiles of the pump (beam axis coincides with the $\rm z$ axis) and probe laser (beam axis along $\hat{\vec{k}}$) beams are depicted in orange and blue, respectively. The pump is modeled as a linearly polarized, pulsed Gaussian beam of waist $w_0$ at ${\rm z}=0$.
The normalized electric $\hat{\vec{e}}_E$ ($\hat{\vec{e}}_e$) and magnetic $\hat{\vec{e}}_B$ ($\hat{\vec{e}}_b$) field vectors of the pump (probe) are perpendicular to its propagation direction $\hat{\vec{e}}_{\rm z}$ ($\hat{\vec{k}}$).
Their orientation is controlled by an angle parameterizing rotations around each beam's propagation axis.
As the x-ray probe is substantially less focused than the pump, in the interaction volume we model it as of constant width;
in the far-field the divergence of the probe needs to be accounted for (cf. main text).
We allow for generic elliptically shaped probe cross-sections: To this end we introduce two different probe beam waists $\{w_1,w_2\}$ associated with two perpendicular transverse directions resembling $\{\hat{\vec{e}}_e,\hat{\vec{e}}_b\}$, but being parameterized by an independent angle.
The vector $\vec{x}_0$ allows for describing a finite impact or spatial displacement of the foci.
We look for signal photons scattered into $\hat{\vec{k}}'$ direction in the far-field. The associated transverse polarization vectors are $\{\pmb{\epsilon}_{(1)},\pmb{\epsilon}_{(2)}\}$.
}
\label{fig:3d}
\end{figure}

Here we assume both laser pulses to be linearly polarized and to feature crossed electric and magnetic fields of the same amplitude profile and oriented mutually perpendicular to the beams' axes.
This assumption is satisfied by Gaussian beams in the paraxial approximation \cite{Siegman,Karbstein:2015cpa}.
Without loss of generality we assume the pump pulse to propagate along the positive $\rm z$ axis
and its electric and magnetic fields to point in $\hat{\vec{e}}_{E}=(\cos\phi,\sin\phi,0)$ and $\hat{\vec{e}}_{B}=(-\sin\phi,\cos\phi,0)$ directions.
The probe may enter from an arbitrary direction $\hat{\vec{k}}=(\cos\varphi\sin\vartheta,-\sin\varphi\sin\vartheta,-\cos\vartheta)$ characterized by the two angles $\varphi$, $\vartheta$.
Its electric and magnetic fields point along $\hat{\vec{e}}_e=\hat{\vec{e}}_{\varphi,\vartheta,\beta}$ and $\hat{\vec{e}}_b=\hat{\vec{e}}_e|_{\beta\to\beta+\frac{\pi}{2}}$, respectively.
In turn, the polarization four vectors of the pump and probe read $\epsilon_{\rm pump}^\mu=(0,\hat{\vec{e}}_E)$ and $\epsilon_{\rm probe}^\mu(\hat{\vec{k}})=(0,\hat{\vec{e}}_e)$.
The pump (probe) polarization is fixed by choosing the angle parameter $\phi$ ($\beta$) accordingly. See Fig.~\ref{fig:3d} for an illustration.
Correspondingly, the combined pump and probe electromagnetic fields constituting $F^{\mu\nu}(x)$ are given by $\vec{E}(x)={\cal E}\hat{\vec{e}}_E+{\frak E}\hat{\vec{e}}_e$ and $\vec{B}(x)={\cal E}\hat{\vec{e}}_B+{\frak E}\hat{\vec{e}}_b$. Here, ${\cal E}\equiv{\cal E}(x)$ and ${\frak E}\equiv{\frak E}(x)$ denote the field profile of the pump and probe laser pulses, respectively. Their explicit expressions will be discussed in Sec.~\ref{sec:specconf}.

Plugging these fields into \Eqref{eq:Sp2} we obtain contributions linear $\sim{\frak E}{\cal E}^2$ and quadratic $\sim{\frak E}^2{\cal E}$ in the probe field.
In the scenario considered by us we generically have ${\frak E}\ll{\cal E}$, such that is sufficient to keep only the terms linear in $\frak E$ here.
To linear order in $\frak E$ (cf. Fig.~\ref{fig:Feyndiag}) this results in
\begin{multline}
{\cal S}_{(p')}(\vec{k}')
 ={\rm i}m^2\frac{\sqrt{\alpha}}{45}\Bigl(\frac{2}{\pi}\Bigr)^{\frac{3}{2}}\sqrt{\rk'}\,
 (1+\cos\vartheta')(1+\cos\vartheta) \bigl[4\cos\gamma'\cos\gamma + 7\sin\gamma'\sin\gamma\bigr]\\
 \times\frac{e{\frak E}_0}{2m^2}\Bigl(\frac{e{\cal E}_0}{2m^2}\Bigr)^2\,{\cal M}\,,
 \label{eq:Sp3}
\end{multline}
where we have introduced the angle parameters $\gamma=\varphi+\beta+\phi$ and $\gamma'=\varphi'+\beta'+\phi$, which encode the entire polarization dependence of the pump, probe and signal photons.
Moreover, we have introduced the peak field strengths of the pump ${\cal E}_0$ and probe ${\frak E}_0$, and have defined
\begin{equation}
 {\cal M}=\int{\rm d}^4x\,{\rm e}^{{\rm i}k'x} \, \frac{{\frak E}(x)}{{\frak E}_0} \Bigl(\frac{{\cal E}(x)}{{\cal E}_0}\Bigr)^2 \,. \label{eq:def=M}
\end{equation}
Note, that for constant fields we would have ${\cal M}\to (2\pi)^4\delta(k')$ and no signal photons are induced.
For a plane wave probe ${\frak E}(x)\sim{\frak E}_0{\rm e}^{-{\rm i}kx}$ traversing a constant pump field we would have ${\cal M}\sim(2\pi)^4\delta(k-k')$, ensuring momentum conservation for the signal photons.

In turn, the differential number of signal photons with polarization $p'$ as defined in \Eqref{eq:Fermi} can be compactly represented as  
\begin{multline}
{\rm d}^3N_{(p')}
 =m^4\frac{{\rm d}^3k'}{(2\pi)^3}\frac{\alpha}{45^2}\,\rk'
 (1+\cos\vartheta')^2(1+\cos\vartheta)^2 \bigl[4\cos\gamma'\cos\gamma + 7\sin\gamma'\sin\gamma\bigr]^2 \\
 \times\Bigl(\frac{e{\frak E}_0}{2m^2}\Bigr)^2\Bigl(\frac{e{\cal E}_0}{2m^2}\Bigr)^4 \Bigl(\frac{2}{\pi}\Bigr)^{3}\,\bigl|{\cal M}\bigr|^2 \,,
 \label{eq:d3N}
\end{multline}
where
${\rm d}^3k'=\rk'^2{\rm dk}'{\rm d}\varphi'{\rm d}\cos\vartheta'$.
The total differential number of signal photons, ${\rm d}^3N=\sum_{p'}{\rm d}N_{(p')}$, is obtained upon summation over the signal photon polarizations.
It can be inferred from \Eqref{eq:d3N} by substituting ${\rm d}^3N_{(p')} \to {\rm d}^3N$ and $[ 4 \cos\gamma'\cos\gamma + 7 \sin\gamma'\sin\gamma ]^2 \to [ 16  + 33 \sin^2\gamma ]$.

In a next step, we aim at determining the number of signal photons polarized perpendicular to the polarization of probe beam $\epsilon_{\rm probe}^\mu(\hat{\vec{k}})$.
We denote the polarization vector of the signal photons $\epsilon^\mu_{(p')}(\hat{\vec{k}}')$ fulfilling this requirement by $\epsilon^\mu_{\perp}(\hat{\vec{k}}')$.
As explained above, these perpendicularly polarized signal photons will constitute the experimental signature of vacuum birefringence.
Ideally the polarization of the probe could then be completely blocked, and the perpendicularly polarized photons reliably detected.
To this end we demand that there is no overlap between the polarization vectors $\epsilon_{\rm probe}^\mu(\hat{\vec{k}})$ and $\epsilon^\mu_{(p')}(\hat{\vec{k}}')$, i.e., require $\epsilon_{(p')\mu}(\hat{\vec{k}}')\epsilon_{\rm probe}^\mu(\hat{\vec{k}})=0$. Solving this equation for $\beta'$ and denoting the solution by $\beta_\perp'$, we obtain
\begin{equation}
 \beta'_\perp=\arctan\biggl\{\frac{\sin\beta\cos\vartheta'\sin(\varphi-\varphi')-\cos\beta[\cos\vartheta\cos\vartheta'\cos(\varphi-\varphi')+\sin\vartheta\sin\vartheta']}{\sin\beta\cos(\varphi-\varphi')+\cos\beta\cos\vartheta\sin(\varphi-\varphi')}\biggr\}
\label{eq:beta'perp}
\end{equation}
and $\epsilon^{\mu}_\perp(\hat{\vec{k}}')=(0,\hat{\vec{e}}_{\varphi',\vartheta',\beta'_\perp})$, which is fully determined by the propagation directions of the probe ($\varphi,\vartheta$) and signal ($\varphi',\vartheta'$) photons, as well as the polarization of the probe ($\beta$).
In turn, the number of signal photons scattered in the perpendicular polarization mode constituting the signature of vacuum birefringence is given by
${\rm d}^3N_{\perp}={\rm d}^3N_{(p')}\big|_{\gamma'\to\gamma'_\perp}$, with ${\rm d}^3N_{(p')}$ as defined in \Eqref{eq:d3N} and  $\gamma'_\perp=\varphi'+\beta'_\perp+\phi$.
For $\vec{k}'\approx\vec{k}$, implying $\varphi'\approx\varphi$ and $\vartheta'\approx\vartheta$, we have $\beta'_\perp\approx (\beta\ {\rm mod}\ \pi)-\frac{\pi}{2}$.

\section{The specific pump and probe field configuration} \label{sec:specconf}

In the following, we specify in detail the macroscopic electromagnetic field profiles invoked to realistically model the high-intensity pump and x-ray probe laser pulses in an actual experiment aiming at the experimental verification of vacuum birefringence.

We assume the high-intensity pump pulse to be well-described by a pulsed Gaussian laser beam in the paraxial approximation.
The corresponding amplitude profile is
\begin{equation}
 {\cal E}(x)={\cal E}_0\,{\rm e}^{-\frac{({\rm z}-t)^2}{(\tau/2)^2}}\frac{w_0}{w({\rm z})} {\rm e}^{-\frac{\rx^2+\ry^2}{w^2({\rm z})}}
\cos\bigl(\Phi(x)\bigr)\,, \label{eq:E(x)}
\end{equation}
with $\Phi(x)=\Omega({\rm z}-t)+\tfrac{\Omega r^2}{2R({\rm z})}-\arctan\bigl(\tfrac{\rm z}{{\rm z}_R}\bigr)$ \cite{Siegman}.
Here ${\cal E}_0$ denotes the peak field strength, $\Omega$ the frequency and $\tau$ the pulse duration;
$w({\rm z})=w_0\sqrt{1+(\frac{\rm z}{{\rm z}_R})^2}$ describes the widening of the beam's transverse extent as a function of $\rm z$,
with $w_0$ the beam's waist size and ${\rm z}_R=\frac{\pi w_0^2}{\lambda}$ its Rayleigh range.
Moreover, $R({\rm z})={\rm z}\bigl[1+(\frac{{\rm z}_R}{\rm z})^2\bigr]$ is the radius of curvature of the wavefronts, and the term $\arctan\bigl(\tfrac{\rm z}{{\rm z}_R}\bigr)$ accounts for the Gouy phase shift.
The square of \Eqref{eq:E(x)} entering \Eqref{eq:Sp3} can be represented as
\begin{equation}
 {\cal E}^2(x)=\frac{1}{4}\,{\cal E}_0^2\,{\rm e}^{-2\frac{({\rm z}-t)^2}{(\tau/2)^2}}
 \biggl[2\Bigl(\frac{w_0}{w({\rm z})}\Bigr)^2{\rm e}^{-2\frac{\rx^2+\ry^2}{w^2({\rm z})}}
 +\sum_{l=\pm1}\frac{1}{(1+il\frac{\rm z}{{\rm z}_R})^2}\,{\rm e}^{-2\frac{\rx^2+\ry^2}{w_0^2(1+il\frac{\rm z}{{\rm z}_R})}}{\rm e}^{il2\Omega({\rm z}-t)}\biggr] .
 \label{eq:E(x)^2}
\end{equation}

In \cite{Karbstein:2015xra} it was found that the terms involving photon exchanges with the pump laser pulse are suppressed by many orders of magnitude (cf. in particular the inlay in Fig.~3 of \cite{Karbstein:2015xra}), making their contributions to the transition amplitude~\eqref{eq:Sp3} practically irrelevant.
Hence, we limit ourselves to the dominant -- $\Omega$ independent -- part of \Eqref{eq:E(x)^2} in the following calculation and use the approximation
\begin{equation}
 {\cal E}^2(x)\approx\frac{1}{2}\,{\cal E}_0^2\,{\rm e}^{-2\frac{({\rm z}-t)^2}{(\tau/2)^2}}
 \Bigl(\frac{w_0}{w({\rm z})}\Bigr)^2{\rm e}^{-2\frac{\rx^2+\ry^2}{w^2({\rm z})}} \,.
 \label{eq:E(x)^2approx}
\end{equation}
Note that this essentially amounts to approximating the square of \Eqref{eq:E(x)} by its envelope, averaging the field's modulation with the laser frequency over one period,
which amounts to the replacement $\cos^2\bigl(\Phi(x)\bigr)\to\frac{1}{2}$.

In the next step we specify the details of the probe pulse.
As the wavelength of the x-ray probe is significantly smaller than the wavelength of an optical high-intensity laser, the probe beam is to be focused only comparatively weakly.
Hence, focusing effects essentially play no role in the interaction volume, and we neglect them in our explicit calculations.
They will nevertheless be relevant for far-field considerations (cf. below).
We parameterize the probe pulse of frequency $\omega$, peak field strength ${\frak E}_0$ and pulse duration $T$ as
\begin{equation}
 {\frak E}(x)={\frak E}_0\,{\rm e}^{-\frac{[\hat{k}(x-x_0)]^2}{(T/2)^2}}\,
 {\rm e}^{-\frac{[\hat{\vec{a}}\cdot(\vec{x}-\vec{x}_0)]^2}{w_1^2}-\frac{[\hat{\vec{b}}\cdot(\vec{x}-\vec{x}_0)]^2}{w_2^2}}
 \cos\bigl(\omega\hat{k}(x-x_0)+\psi_0\bigr) . \label{eq:Eprobe}
\end{equation}
Here $\hat{k}^\mu=(1,\hat{\vec{k}})$ is the probe's normalized four wave-vector ($k^\mu=\omega\hat{k}^\mu$), $\psi_0$ is a constant phase,
and $x_0^\mu=(t_0,\vec{x}_0)=(t_0,\rx_0,\ry_0,\rz_0)$ denotes a spatio-temporal offset relative to the reference point, which is chosen as the focus of the pump laser pulse at $x^\mu=(0,\vec{0})$.
It allows for a time delay and an impact parameter under which the pump is hit by the probe pulse.
Moreover, we account for the possibility of generically oriented, elliptically shaped probe-beam cross-sections.
To this end we introduce two different beam waist parameters $\{w_1,w_2\}$ associated with two perpendicular directions $\{\hat{\vec{a}},\hat{\vec{b}}\}$ transverse to the beam's propagation direction $\hat{\vec{k}}$.
We choose them as $\hat{\vec{a}}=\hat{\vec{e}}_{\varphi,\vartheta,\delta_0}$ and $\hat{\vec{b}}=\hat{\vec{e}}_{\varphi,\vartheta,\delta_0+\frac{\pi}{2}}$, such that the precise orientation of the ellipse describing the beam's cross-section can be adjusted by the angle $\delta_0$.
In turn, the radius (waist) of the probe beam depends on the angle $\delta$ parameterizing a rotation around $\hat{\vec k}$ and reads\footnote{This is the $1/{\rm e}$ radius (divergence) for the field strength, or equivalently the $1/{\rm e}^2$ radius (divergence) for the intensity.}
${\rm w}(\delta)=\sqrt{w_1^2\cos^2(\delta-\delta_0)+w_2^2\sin^2(\delta-\delta_0)}$; as to be expected, the $\delta$ dependence drops out in the circular limit where $w_1=w_2$.

Assuming the probe to be well-described as a focused Gaussian beam, its divergence in the far field also depends on $\delta$ and is given by$^1$ $\theta(\delta)\simeq\frac{2}{\omega{\rm w}(\delta)}\ll1$.
The number of probe photons $N$ is proportional to the probe's electric field squared, i.e., $N\sim\mathfrak{E}^2$ (cf. also Sec.~\ref{sec:predictions} below).
Hence, we have $\frac{{\rm d}^2N}{{\rm d}\delta\,{\rm d}\!\cos\zeta}\simeq \tilde N\,{\rm e}^{-2(\frac{\zeta}{\theta(\delta)})^2}$, where the polar angle $\zeta$, measured with respect to the probe's beam axis pointing in $\hat{\vec k}$ direction, spans the divergence of the probe beam, and $\tilde N$ is an amplitude.
In order to determine the amplitude $\tilde N$ explicitly we integrate this equation over the angles: Because of $\theta(\delta)\ll1$ the exponential function ensures that the integration over $\zeta$ receives its main contribution from small values of $\zeta$.
In turn, we can formally extend the upper limit of the $\zeta$ integral to infinity and approximate $\sin\zeta\simeq\zeta$, such that $\int_0^{2\pi}{\rm d}\delta\int_0^{\gg\theta(\delta)}{\rm d}\zeta\sin\zeta\to\int_0^{2\pi}{\rm d}\delta\int_0^\infty{\rm d}\zeta\,\zeta$.
Performing the latter integrations, we infer $\tilde N\simeq\frac{\omega^2w_1w_2}{2\pi}N$.

\section{Results}\label{sec:results}

Plugging Eqs.~\eqref{eq:E(x)^2approx} and \eqref{eq:Eprobe} into \Eqref{eq:def=M}, the integrations over $t$, $\rx$ and $\ry$ can be performed explicitly.
They correspond to elementary Gaussian integrals. However, due to the many parameters accounted for here, the resulting expressions are typically rather unhandy and lengthy.
Finding a convenient representation actually turns out to be quite challenging.
Defining several auxiliary parameters and functions and
after some tedious but straightforward manipulations we have nevertheless discovered a rather compact representation of \Eqref{eq:def=M} accounting for the above pump and probe pulse profiles.

First of all, it is convenient to decompose vectors into components parallel and perpendicular to the pump's propagation direction $\hat{\vec{z}}$,
such that, e.g., $\vec{k}_\perp=\vec{k}-(\vec{k}\cdot\hat{\vec{z}})\hat{\vec{z}}$. Moreover, we introduce $\hat{a}^\mu=(0,\hat{\vec{a}})$ and $\hat{b}^\mu=(0,\hat{\vec{b}})$.
We then define the two four-vectors
\begin{gather}
 \mu^\mu=2\biggl(\frac{\hat{\vec{a}}\cdot\vec{x}_0}{w_1^2}\hat{a}^\mu+\frac{\hat{\vec{b}}\cdot\vec{x}_0}{w_2^2}\hat{b}^\mu+4\frac{(\hat{k}x_0)}{T^2}\hat{k}^\mu\biggr) , \\
 \nu^\mu=-2\biggl(\frac{\hat{a}_3}{w_1^2}\hat{a}^\mu+\frac{\hat{b}_3}{w_2^2}\hat{b}^\mu+4\frac{\hat{k}_3}{T^2}\hat{k}^\mu\biggr) ,
\end{gather}
and the functions
\begin{equation}
 j(\vec{c},\vec{d})=\frac{(\hat{\vec{a}}_{\perp}\times\vec{c}_\perp)\cdot(\hat{\vec{a}}_{\perp}\times\vec{d}_\perp)}{w_1^2}
 +\frac{(\hat{\vec{b}}_{\perp}\times\vec{c}_\perp)\cdot(\hat{\vec{b}}_{\perp}\times\vec{d}_\perp)}{w_2^2}+2\frac{\vec{c}_\perp\cdot\vec{d}_\perp}{w^2({\rm z})} \,, \label{eq:j}
\end{equation}
\begin{equation}
 \bar{j}(\vec{c},\vec{d})=j(\vec{c},\vec{d})+4\frac{(\hat{\vec{k}}_\perp\times\vec{c}_\perp)\cdot(\hat{\vec{k}}_\perp\times\vec{d}_\perp)}{T^2} \,, \label{eq:barj}
\end{equation}
which attribute a scalar quantity to any given vectors $\vec{c}$ and $\vec{d}$.
In addition, we make use of the abbreviations $g=\frac{8}{T^2}j(\hat{\vec{k}},\hat{\vec{k}})$ and
\begin{equation}
 f = 2\frac{(\hat{\vec{a}}_\perp\times\hat{\vec{b}}_\perp)^2}{(w_1w_2)^2}
 + \frac{4}{w^2({\rm z})}\Bigl(\frac{\hat{\vec{a}}_\perp^2}{w_1^2}+\frac{\hat{\vec{b}}_\perp^2}{w_2^2}+\frac{2}{w^2(\rz)}\Bigr)
 + g \,. \label{eq:f}
\end{equation}
Note that $j(\hat{\vec{k}},\hat{\vec{k}})\geq\frac{2}{w^2(\rz)}$. In turn, the ratio $\frac{g}{f}$ is constrained by $\frac{8}{T^2}\frac{2}{w^2(\rz)}\frac{1}{f}\leq\frac{g}{f}\leq1$.

\subsection{Generic case}

With the above definitions, our result for $\cal M$ can then be expressed as
\begin{multline}
 {\cal M} = \Bigl(\frac{\pi}{4}\Bigr)^{\frac{3}{2}}\,{\rm e}^{-\frac{(\mu x_0)}{2}}
 \sum_{q=\pm1}
 \int{\rm dz}\,\Bigl(\frac{w_0}{w({\rm z})}\Bigr)^2\frac{\tau}{\sqrt{f\bigl[1+\frac{1}{2}(\frac{\tau}{T})^2(1-\frac{g}{f})\bigr]}}\, 
 {\rm e}^{\frac{\frac{2}{T^2}(\frac{\tau}{T})^2\bigl(\frac{j(\pmb\mu,\hat{\vec{k}})}{f}-(\hat{k}x_0)\bigr)^2}{1+\frac{1}{2}(\frac{\tau}{T})^2(1-\frac{g}{f})}+\frac{\bar{j}(\pmb\mu,\pmb\mu)}{2f}}\\
  \times
 {\rm e}^{-\frac{1}{8}\frac{\frac{(T\omega)^2}{2}\bigl[\frac{1}{2}(\frac{\tau}{T})^2(1-\frac{g}{f})+\frac{g}{f}\bigr]-T^2\omega\rk'q\bigl[\frac{1}{2}(\frac{\tau}{T})^2(1-\frac{g}{f})+\frac{8}{T^2}\frac{j(\hat{\vec{k}}',\hat{\vec{k}})}{f}\bigr]+\frac{(\tau\rk')^2}{4} \bigl(1-\frac{8}{T^2}\frac{j(\hat{\vec{k}}',\hat{\vec{k}})}{f}\bigr)^2}{1+\frac{1}{2}(\frac{\tau}{T})^2(1-\frac{g}{f})}
 -\rk'^2\frac{\bar{j}(\hat{\vec{k}}',\hat{\vec{k}}')}{2f}}\\
 \times{\rm e}^{-\Bigl\{\frac{8}{\tau^2}-\frac{2}{\tau^2}\frac{\bigl[2+(\frac{\tau}{T})^2\bigl(\frac{j(\pmb\nu,\hat{\vec{k}})}{f}+\hat{k}_3\bigr)\bigr]^2}{1+\frac{1}{2}(\frac{\tau}{T})^2(1-\frac{g}{f})}
 -\frac{\bar{j}(\pmb\nu,\pmb\nu)}{2f}-\frac{\nu_3}{2}\Bigr\}\rz^2}
 {\rm e}^{\Bigl\{\frac{8}{T^2}\frac{\bigl[1+\frac{1}{2}(\frac{\tau}{T})^2\bigl(\frac{j(\pmb\nu,\hat{\vec{k}})}{f}+\hat{k}_3\bigr)\bigr]\bigl(\frac{j(\pmb\mu,\hat{\vec{k}})}{f}-(\hat{k}x_0)\bigr)}{1+\frac{1}{2}(\frac{\tau}{T})^2(1-\frac{g}{f})}
 +\frac{\bar{j}(\pmb\mu,\pmb\nu)}{f}+\mu_3\Bigr\}\rz}\\
 \times{\rm e}^{{\rm i}\Bigl\{
 \tfrac{q\omega(1-\frac{g}{f})-\rk'\bigl(1-\frac{8}{T^2}\frac{j(\hat{\vec{k}}',\hat{\vec{k}})}{f}\bigr)-\bigl[q\omega+\frac{1}{2}(\frac{\tau}{T})^2\rk'\bigl(1-\frac{8}{T^2}\frac{j(\hat{\vec{k}}',\hat{\vec{k}})}{f}\bigr)\bigr]\bigl(\frac{j(\pmb\nu,\hat{\vec{k}})}{f}+\hat{k}_3\bigr)}{1+\frac{1}{2}(\frac{\tau}{T})^2(1-\frac{g}{f})}
 +\rk'\bigl(\frac{\bar{j}(\pmb\nu,\hat{\vec{k}}')}{f}+\hat{k}'_3\bigr)\Bigr\}\rz}\\
 \times{\rm e}^{{\rm i}\Bigl\{\rk'\frac{\bar{j}(\pmb\mu,\hat{\vec{k}}')}{f}-\tfrac{\bigl[q\omega+\frac{1}{2}(\frac{\tau}{T})^2\rk'\bigl(1-\frac{8}{T^2}\frac{j(\hat{\vec{k}}',\hat{\vec{k}})}{f}\bigr)\bigr]\bigl(\frac{j(\pmb\mu,\hat{\vec{k}})}{f}-(\hat{k}x_0)\bigr)}{1+\frac{1}{2}(\frac{\tau}{T})^2(1-\frac{g}{f})}-q\psi_0\Bigr\}} .
 \label{eq:M}
\end{multline}
The $\rz$ integration in \Eqref{eq:M} can in general not be performed analytically; recall the implicit $w(\rz)$ dependence of \Eqref{eq:M} encoded in the functions \eqref{eq:j}-\eqref{eq:f}.
Also note that the integrand in \Eqref{eq:M} depends on the modulus of the signal photon momentum $\rk'$ only via linear and quadratic terms in the exponential.
Setting all offset parameters $x_0^\mu$ equal to zero, \Eqref{eq:M} simplifies significantly:
This choice implies $\mu^\mu=0$, such that all the functions $j(.,.)$ and $\bar j(.,.)$ with at least one of their arguments being $\pmb\mu$ vanish.

As all the functions and parameters in \Eqref{eq:M} are purely real-valued \Eqref{eq:M} can be easily decomposed into its real and imaginary parts, employing ${\rm e}^{\ri\chi}=\cos\chi+\ri\sin\chi$.
This decomposition is of relevance when aiming at the determination of the modulus squared of \Eqref{eq:M},
$|{\cal M}|^2=\Re^2({\cal M})+\Im^2({\cal M})$, entering the expression for the differential number of signal photons \Eqref{eq:d3N}.

Due to its substantially lower frequency and scales of variation, the pump pulse cannot affect x-ray frequencies and momenta significantly.
Correspondingly, the signal photons are essentially induced in the probe's propagation direction, and we have $\vec{k}'\approx\vec{k}$.
This in particular implies $\rk'\approx\omega$, $\frac{8}{T^2}j(\hat{\vec{k}}',\hat{\vec{k}})\approx g$ and $\bigl(\frac{\bar{j}(\pmb\nu,\hat{\vec{k}}')}{f}+\hat{k}_3'\bigr)\approx\bigl(\frac{j(\pmb\nu,\hat{\vec{k}})}{f}+\hat{k}_3\bigr)$.
Employing these approximations, it is easy to see that  in \Eqref{eq:M} the $q=-1$ contribution is negligible in comparison to the $q=+1$ one:
First of all, the $q$ dependence of the expression in the second line of \Eqref{eq:M} is in terms of
$\approx\exp\bigl\{\frac{q}{2}(\frac{T\omega}{2})^2[\frac{1}{2}(\frac{\tau}{T})^2(1-\frac{g}{f})+\frac{g}{f}]/[1+\frac{1}{2}(\frac{\tau}{T})^2(1-\frac{g}{f})]\bigr\}$, implying an exponential suppression of the $q=-1$ term relatively to the $q=+1$ term.
Secondly, due to the term in the fourth line of \Eqref{eq:M}, which becomes
$\approx\exp\bigl\{{\rm i}(q-1)\omega\rz[1-\frac{g}{f}-\frac{{j}(\pmb\nu,\hat{\vec{k}})}{f}-\hat{k}_3]/[1+\frac{1}{2}(\frac{\tau}{T})^2(1-\frac{g}{f})]\bigr\}$,
the $q=-1$ contribution exhibits a rapid oscillation with $\omega\rz$, rendering the $\rz$ integration practically zero, while no such oscillation is encountered for the $q=+1$ contribution.
For these reasons -- and to avoid unnecessary computational efforts -- in the explicit analyses performed below for given experimental parameters we limit ourselves to the $q=+1$ contribution.

Finally note that when replacing $w(\rz)\to w_0$ in \Eqref{eq:M}, i.e., formally turning to a pump with an infinite Rayleigh length, the $\rz$ integration is of Gaussian type and can be performed easily.
Upon insertion into \Eqref{eq:d3N}, in this limit even the $\rk'$ integration can be performed explicitly without difficulty, resulting in an analytic expression for $\frac{{\rm d}^2N_{(p')}}{{\rm d}\varphi'{\rm d}\!\cos\vartheta'}$. 

An approximation of \Eqref{eq:M} which allows for an analytic evaluation of the $\rm z$ integration can be found in Appendix~\ref{app:Appendix}. 
This might be useful to guide optimization studies necessitating the variation of many parameters, e.g., different offset parameters and collision angles:
Sticking to this approximation, the $\rz$ integration does not have to be performed numerically, which reduces the numerical efforts in the determination of the number of perpendicularly polarized signal photons significantly.
In the present work we will only provide explicit results for idealized conditions, i.e., $x_0^\mu=0$ and counter-propagating pump and probe laser pulses (cf. Sec.~\ref{sec:predictions} below). For this case the exact results are readily integrated numerically with standard tools.

\subsection{Counter-propagation geometry} \label{subsec:counterprop}

Sticking to a counter-propagation geometry for the pump and probe laser pulses, i.e., $\hat{k}^\mu=(1,-\hat{\vec{e}}_{\rm z})\ \leftrightarrow\ \vartheta=0$,
and without loss of generality adopting the choice of $\varphi=0$,
we have $\epsilon^\mu_{\rm probe}(\hat{\vec{k}})=(0,-\cos\beta,\sin\beta,0)$.
Moreover, as the signal photons are predominantly emitted in forward direction,
\Eqref{eq:beta'perp} with $\vartheta=\varphi=0$ can be approximated as $\beta'_\perp=[(\beta-\varphi')\ {\rm mod}\ \pi]-\frac{\pi}{2}+{\cal O}(\vartheta'^2)$.
In turn $[4\cos\gamma'_\perp\cos\gamma + 7\sin\gamma'_\perp\sin\gamma]^2\to\frac{9}{4}\sin^2(2\gamma)$, which implies that the number of perpendicularly polarized signal photons can be maximized by choosing $\gamma=\frac{\pi}{4}(1+2n)\ \leftrightarrow\ \beta=\frac{\pi}{4}(1+2n)-\phi$, with $n\in\mathbb{Z}$. Note that for this choice the polarization vector of the probe forms an angle of $\frac{\pi}{4}$ with both the electric and magnetic field vectors of the pump.
The differential number of perpendicularly polarized signal photons in this particular limit can be inferred from the above equations and reads
\begin{equation}
{\rm d}^3N_{\perp}
 \approx m^4\frac{{\rm d}^3k'}{(2\pi)^3}\frac{\alpha}{15^2}\,\rk'
 (1+\cos\vartheta')^2
 \Bigl(\frac{e{\frak E}_0}{2m^2}\Bigr)^2\Bigl(\frac{e{\cal E}_0}{2m^2}\Bigr)^4 \Bigl(\frac{2}{\pi}\Bigr)^{3}\,\bigl|{\cal M}_0\bigr|^2 \,,
 \label{eq:d3Nperpcp}
\end{equation}
with
\begin{multline}
 {\cal M}_0 = 
\Bigl(\frac{\pi}{4}\Bigr)^{\frac{3}{2}}\frac{\tau}{\sqrt{1+\frac{1}{2}(\frac{\tau}{T})^2}}
\sum_{q=\pm1} {\rm e}^{-\frac{2}{1+\frac{1}{2}(\frac{\tau}{T})^2}\bigl[\tau^2(\frac{q\omega-\rk'}{8})^2+2(\frac{\rz_0+t_0}{T})^2\bigr]}
\int{\rm dz}\,\Bigl(\frac{w_0}{w({\rm z})}\Bigr)^2 \frac{1}{\sqrt{f_0}}\, \\
\times {\rm e}^{-\frac{4}{w^2(\rz)f_0}\bigl[\bigl(\frac{1}{w_2^2}+\frac{2}{w^2(\rz)}\bigr)(\frac{\hat{\vec{a}}\cdot\vec{x}_0}{w_1})^2
+\bigl(\frac{1}{w_1^2}+\frac{2}{w^2(\rz)}\bigr)(\frac{\hat{\vec{b}}\cdot\vec{x}_0}{w_2})^2
\bigr]-\frac{(\rk'\sin\vartheta')^2}{2f_0}\bigl[\bigl(\frac{{\rm w}(\varphi')}{w_1w_2}\bigr)^2+\frac{2}{w^2(\rz)}\bigr]}  \\
 \times{\rm e}^{-(\frac{4}{T})^2 \frac{\rz^2-(\rz_0+t_0)\,\rz}{1+\frac{1}{2}(\frac{\tau}{T})^2}}\,
 {\rm e}^{{\rm i}\bigl[\frac{2(q\omega-\rk')}{1+\frac{1}{2}(\frac{\tau}{T})^2}+\rk'(1-\cos\vartheta')\bigr]\rz}   \\
 \times{\rm e}^{-{\rm i}\bigl\{\frac{(q\omega-\rk')(\rz_0+t_0)}{1+\frac{1}{2}(\frac{\tau}{T})^2}+\frac{2\rk'\sin\vartheta'}{f_0}\bigl[\frac{\cos(\varphi'-\delta_0)}{w_1}\bigl(\frac{1}{w_2^2}+\frac{2}{w^2(\rz)}\bigr)\frac{\hat{\vec{a}}\cdot\vec{x}_0}{w_1}
+\frac{\sin(\varphi'-\delta_0)}{w_2}\bigl(\frac{1}{w_1^2}+\frac{2}{w^2(\rz)}\bigr)\frac{\hat{\vec{b}}\cdot\vec{x}_0}{w_2}\bigr]+\rk'(\rz_0+t_0)+q\psi_0\bigr\}} .
 \label{eq:Mcp}
\end{multline}
Here, we made use of the shorthand notations ${\rm w}(\varphi')=\sqrt{w_1^2\cos^2(\varphi'-\delta_0)+w_2^2\sin^2(\varphi'-\delta_0)}$, ${\cal M}_0={\cal M}|_{\vartheta=0}$ and
\begin{equation}
 f_0 = f|_{\vartheta=0} = \frac{2}{(w_1w_2)^2} + \frac{4}{w^2({\rm z})}\Bigl(\frac{1}{w_1^2}+\frac{1}{w_2^2}+\frac{2}{w^2(\rz)}\Bigr) .  \label{eq:fcp}
\end{equation}
In the considered limit we have
$\hat{\vec{a}}=\hat{\vec{e}}_{0,0,\delta_0}$ and $\hat{\vec{b}}=\hat{\vec{e}}_{0,0,\delta_0+\frac{\pi}{2}}$.
From \Eqref{eq:Mcp} it is particularly obvious that the $q=-1$ contribution is substantially suppressed in comparison to the one with $q=+1$.
Moreover, note that for $\hat{\vec{a}}\cdot\vec{x}_0=\hat{\vec{b}}\cdot\vec{x}_0=0$, which amounts to zero impact parameter, \Eqref{eq:Mcp} depends on the azimuthal angle $\varphi'$ only via the probe waist ${\rm w}(\varphi')$.
If in addition $w_1=w_2$, i.e., for circular cross-sections of the probe beam, ${\cal M}_0$ becomes independent of $\varphi'$ and the orientation of $\hat{\vec{a}}$ and $\hat{\vec{b}}$ controlled by the angle parameter $\delta_0$.
Also note that even for $w_1\neq w_2$ the total (integrated) numbers of attainable signal photons are always independent of the specific choice for $\delta_0$. In other words, the orientation of the probe's cross-section controlled by $\{\hat{\vec{a}},\hat{\vec{b}}\}$ relative to the probe's field vectors $\{\hat{\vec{e}}_e,\hat{\vec{e}}_b\}$ does not affect the total numbers of signal photons.

\section{Predictions for experiments}\label{sec:predictions}

In the following we adopt the parameters of a state-of-the-art high-intensity laser and a FEL facility to provide for realistic estimates of the attainable numbers of perpendicularly polarized signal photons.
The pump pulse is provided by an $1$PW class laser of optical or near-infrared frequency (pulse energy $W=30$J, pulse duration $\tau=30$fs) focused to $w_0=1\mu$m.
As the frequency of the pump does not enter our expression for the differential number of signal photons, but only manifests itself in neglected subleading contributions (cf. Sec.~\ref{sec:specconf} above), we do not need to specify it here.
Correspondingly, the peak intensity of the pump is ${\cal I}_0={\cal E}_0^2=2\frac{0.87\,W}{\pi w_0^2 \tau}$, where we account for the fact that the effective focus volume delimited by $w_0$ and $\tau$
contains a factor of ${\rm erf}^3(\sqrt{2})\approx0.87$ of the total pulse energy; ${\rm erf}(.)$ is the error function.
The x-ray probe is assumed to deliver $N$ photons of energy $\omega=12914$eV per pulse of duration $T=30$fs; aiming at explicit numerical estimates for the numbers of signal photons to be detected experimentally we assume $N=10^{12}$.
Precisely for this photon energy the presently most sensitive x-ray polarimeter \cite{Uschmann:2014} was benchmarked.
The polarization purity of x-rays of this energy can be measured to the level of ${\cal P}=5.7\cdot10^{-10}$.
From the above parameters we infer the peak intensity of the probe, which is given by $\frak{I}_0=\frak{E}_0^2=2\frac{0.87\,N \omega}{\pi w_1w_2 T}$.
For the moment, we leave the probe beam widths $w_1$ and $w_2$ unspecified. Below we will discuss the effect of different choices of these parameters. 
As the differential numbers of signal photons~\eqref{eq:d3N} and \eqref{eq:d3Nperpcp} scale as ${\cal E}_0^4\sim{\cal I}_0^2\sim W^2$ our results for ${\rm d}^3N_\perp$ can straightforwardly be rescaled as $(\frac{W[{\rm J}]}{30})^2$ to any other pump laser energy $W$.
Analogously, due to the scaling with $\frak{E}_0^2\sim\frak{I}_0\sim N$, the dependence of ${\rm d}^3N_\perp$ on the number of x-ray probe photons is only in terms of an overall factor of $N$.

It is particularly interesting to compare our new results with those previously obtained in \cite{Karbstein:2015xra}.
As noted above, this study focused on a counter-propagation geometry, but was manifestly limited to certain special cases, namely probe beams either significantly narrower or wider than the pump beam.
More precisely, the considered cases were exactly those where the details of the transverse profile of the probe should arguably not have any significant effect on the estimated numbers of attainable signal photons.
The fact that we can now consider arbitrary probe beam widths calls for a critical reassessment of these estimates.

The results for counter-propagating pump and probe laser pulses follow from \Eqref{eq:d3Nperpcp} upon insertion of \Eqref{eq:Mcp} involving a numeric evaluation of the $\rz$ integral.
The maximum number of signal photons is obtained for vanishing offset parameters, $x_0^\mu=0$, ensuring that the pump and probe beams have an optimal overlap.
Note that the only offset parameter accounted for in \cite{Karbstein:2015xra} was the time delay $t_0$.

Our main focus is on the differential number of signal photons $\frac{{\rm d}^2N_\perp}{{\rm d}\varphi'\,{\rm d}\!\cos\vartheta'}\equiv\int_0^\infty{\rm dk}'\,\rk'^2\,\frac{{\rm d}^3N}{{\rm d}^3k'}$.
This quantity generically exhibits a maximum at $\vartheta'=\vartheta=0$, implying that most of the signal photons are emitted exactly in forward direction, and falls off rapidly with $\vartheta'$.
Particularly for asymmetric probe beam cross-sections with $w_1\neq w_2$ the corresponding signal photon distribution is also asymmetric, resulting in differently pronounced decays with $\vartheta'$ as a function of the azimuthal angle $\varphi'$.
It is instructive to compare $\frac{{\rm d}^2N_\perp}{{\rm d}\varphi'{\rm d}\cos\vartheta'}$ to the far-field direction distribution of the probe photons having traversed the interaction volume practically unaffected.
For the counter-propagation geometry considered here, the angles $\{\delta,\zeta\}$ introduced in Sec.~\ref{sec:specconf} above can be identified with $\{\varphi',\vartheta'\}$, such that this distribution is given by
$\frac{{\rm d}^2N}{{\rm d}\varphi'\,{\rm d}\!\cos\vartheta'}\simeq \frac{\omega^2w_1w_2}{2\pi}N\,{\rm e}^{-2(\frac{\vartheta'}{\theta(\varphi')})^2}$
with $\theta(\varphi')=\frac{\theta_1\theta_2}{\sqrt{\theta_2^2\cos^2(\varphi'-\delta_0)+\theta_1^2\sin^2(\varphi'-\delta_0)}}$ and $\theta_i=\frac{2}{\omega w_i}$.
We denote the number of perpendicularly polarized signal photons emitted outside the probe beam divergence by $N_{\perp>\theta}$.

\begin{table}
\begin{tabular}{|c|c||c|c|c|c|c|c|c|c|}
 \hline
  $\frac{w_1}{w_0}$ & $\frac{w_2}{w_0}$ & $\frac{N_\perp}{N}$  & $\theta_1[\mu{\rm rad}]$ & $\theta_2[\mu{\rm rad}]$ & $\frac{N_{\perp>\theta}}{N_\perp}$ & $\sigma_1[\mu{\rm rad}]$ & $\sigma_2[\mu{\rm rad}]$ & $\frac{N_{\perp>\sigma}}{N_\perp}\phantom{\Big.^{|}}$\hspace*{-1mm} & $\frac{N_{\perp>\sigma}}{\rm h}$ \\
 \hline
 \hline
  $\frac{1}{10}$ & $\frac{1}{10}$ & $6.12\cdot10^{-13}$  & $306.07$ & $306.07$ & $14.0\%$ & $4293.20$ & $4293.20$ & $0.0\%$ & $0.00$ \\
  $\frac{1}{3}$ & $\frac{1}{3}$ & $5.19\cdot10^{-13}$ & $91.82$ & $91.82$ & $18.9\%$ & $422.46$ & $422.46$ & $0.0\%$ & $0.00$ \\
  $1$ & $1$ & $2.23\cdot10^{-13}$ & $30.61$ & $30.61$ & $48.7\%$ & $80.42$ & $80.42$ & $0.7\%$ & $5.86$ \\
  $3$ & $3$ & $3.64\cdot10^{-14}$ & $10.20$ & $10.20$ & $88.8\%$ & $26.29$ & $26.29$ & $45.6\%$ & $59.64$ \\
  $3$ & $\frac{1}{10}$ & $1.49\cdot10^{-13}$ & $10.20$ & $306.07$ & $68.5\%$ & $23.14$ & $5129.38$ & $27.2\%$ & $145.74$ \\
  $3$ & $\frac{1}{3}$ & $1.37\cdot10^{-13}$  & $10.20$ & $91.82$ & $69.5\%$ & $23.33$ & $494.22$ & $27.0\%$ & $133.35$ \\
  $3$ & $1$ & $8.99\cdot10^{-14}$  & $10.20$ & $30.61$ & $76.0\%$ & $24.31$ & $88.24$ & $27.1\%$ & $87.78$ \\
 \hline
 \end{tabular}
 \caption{Attainable numbers of signal photons for the ideal case scenario of zero impact parameter and time delay between the pump and probe pulses ($x_0^\mu=0$).
 We present results for different probe beam cross sections controlled by the two independent waists $w_i$, measured in units of the pump waist $w_0$;
 the associated divergences are $\theta_i\sim\frac{1}{w_i}$. The probe pulse (duration $\tau=30{\rm fs}$) comprises $N$ photons of energy $\omega=12914{\rm eV}$.
 The high-intensity pump is assumed to be a $1{\rm PW}$ system delivering optical or near-infrared frequency pulses of energy $W=30{\rm J}$ and duration $T=30{\rm fs}$ focused to a waist of $w_0=1\mu{\rm m}$. 
 $N_{\perp}$ ($N_{\perp>\theta}$) is the total number of perpendicularly polarized signal photons (emitted outside the divergence $\theta(\varphi')$ of the probe beam),
and $N_{\perp>\sigma}$ denotes the number of perpendicularly polarized signal photons fulfilling $\vartheta'\geq\sigma(\varphi')$ with $\sigma_i=\sigma(\delta_0+\frac{\pi}{2}\delta_{i2})$ that could be detected with state-of-the-art technology.
 The values provided in the last column, are for $N=10^{12}$ probe photons per pulse and a repetition rate of $1{\rm Hz}$ (cf. main text).}
\label{tab:data}
\end{table}

In Tab.~\ref{tab:data} we provide explicit results for various probe beam cross-sections parameterized by the two beam waists $w_1$ and $w_2$ measured in units of the pump waist $w_0$.
As to be expected, for vanishing offset parameters, $x_0^\mu=0$, the total number of polarization-flipped photons $N_\perp$ is maximal for minimal probe beam cross-sections.
In this case all probe photons propagate close to the optical axis of the pump where the pump field strength triggering the effect is maximal. 
However, another well-known effect is that the divergence of the probe beam in the far-field scales as $\theta\sim\frac{1}{{\rm w}}$, or equivalently $\theta_i\sim\frac{1}{w_i}$ (cf. above).
This implies that for smaller probe waists, in the far-field the probe photons having traversed the interaction volume are distributed over a larger polar angle interval.
If we ask for the fraction of signal photons emitted outside the divergence of the probe, $N_{\perp>\theta}/N_\perp$, we find that also the polar angle spread of the signal photons increases with decreasing probe beam cross-sections.
In fact the fraction of signal photons emitted outside the divergence of the probe beam increases with the probe beam cross-section.
This implies that in the far-field 
the decrease of the differential number of signal photons $\frac{{\rm d}^2N}{{\rm d}\varphi'\,{\rm d}\!\cos\vartheta'}$ with $\vartheta'$ differs significantly from that of the differential number of probe photons $\frac{{\rm d}^2N_\perp}{{\rm d}\varphi'\,{\rm d}\!\cos\vartheta'}$ having traversed the interaction region unaffected:
Generically, the former decreases substantially faster with $\vartheta'$ than the latter.

\begin{figure}
\center
\includegraphics[width=0.85\textwidth]{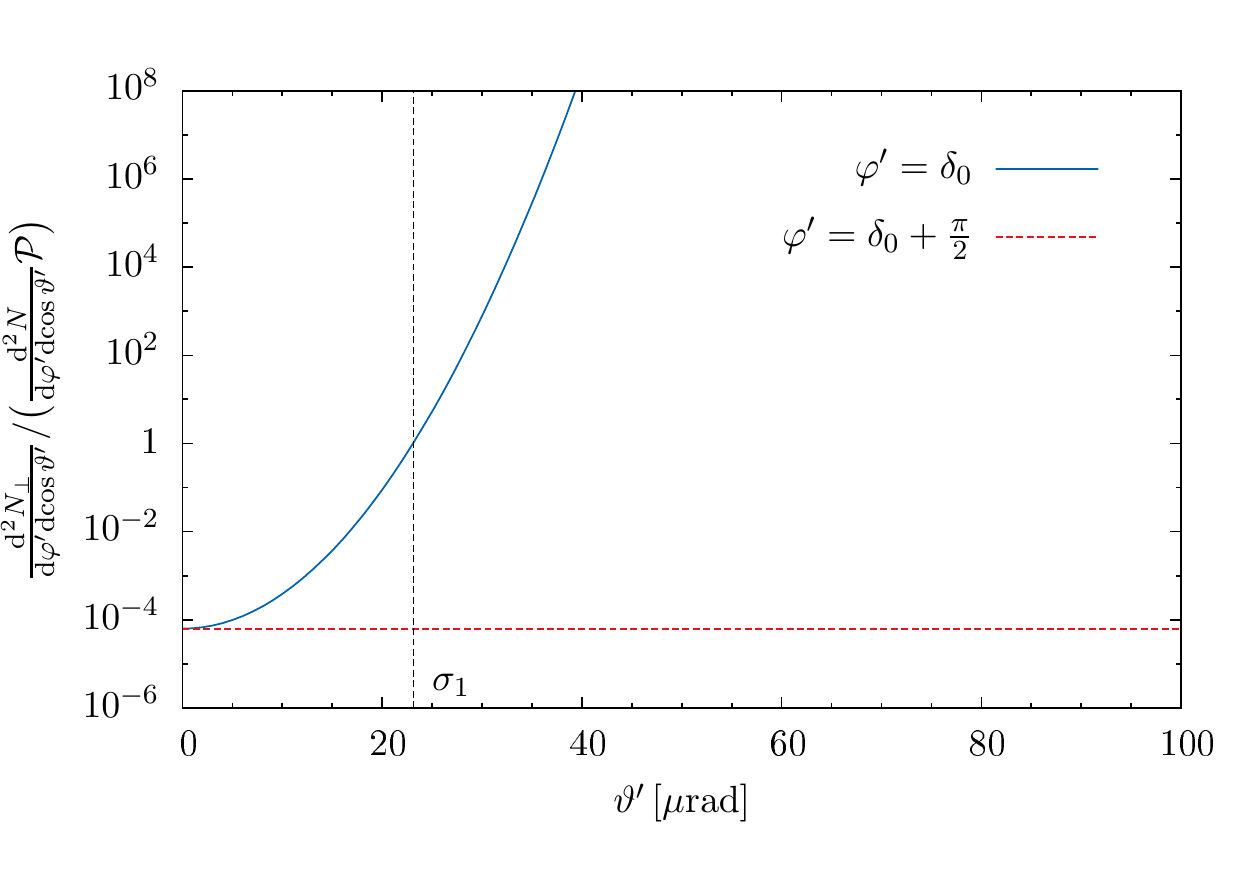}
\caption{Plot of the ratio $\frac{{\rm d}^2N_\perp}{{\rm d}\varphi'\,{\rm d}\!\cos\vartheta'}/\big(\frac{{\rm d}^2N}{{\rm d}\varphi'\,{\rm d}\!\cos\vartheta'}{\cal P}\bigr)$
as a function of the polar angle $\vartheta'$ for a probe of asymmetric cross-section; waist parameters $\frac{w_1}{w_0}=3$ for $\varphi'=\delta_0$, and $\frac{w_2}{w_0}=\frac{1}{10}$ for $\varphi'=\delta_0+\frac{\pi}{2}$. 
The blue (solid) curve for $\varphi'=\delta_0$ becomes unity at $\vartheta'=\sigma_1=23.12\mu$rad. In the depicted angle range, the red (dashed) curve for $\varphi'=\delta_0+\frac{\pi}{2}$
looks consistent with a straight line. In fact, it slowly bents upwards to surpass unity for $\vartheta'=\sigma_2=5129.38\mu$rad (cf. also Tab.~\ref{tab:data}).
The two curves depicted here are continuously related as a function of the azimuthal angle $\varphi'$ (cf. main text).}
\label{fig:d2Nperpbyd2NbyP}
\end{figure}

In a second step we now ask for the number of perpendicularly polarized signal photons that could be measured with state-of-the-art technology.
Our criterion to judge if signal photons emitted in a given solid angle element ${\rm d}\varphi'{\rm d}\!\cos\vartheta'$ can be reliably measured with state-of-the-art technology is as follows: If the ratio $\frac{{\rm d}^2N_\perp}{{\rm d}\varphi'\,{\rm d}\!\cos\vartheta'}/\frac{{\rm d}^2N}{{\rm d}\varphi'\,{\rm d}\!\cos\vartheta'}$ is larger than the polarization purity record $\cal P$ of the presently best x-ray spectrometer \cite{Uschmann:2014}, we can discern the signal photons from the background.
Note that even if this condition is not met for $\vartheta'=0$, it will eventually be fulfilled for large enough values of $\vartheta'$.
Of course, in addition the fraction of signal photons emitted at these angles needs to be large enough to allow for sufficient statistics.
To this end we introduce the two angle parameters $\sigma_i$ determined by means of the following implicit condition
$\frac{{\rm d}^2N_\perp}{{\rm d}\varphi'\,{\rm d}\!\cos\vartheta'}/\big(\frac{{\rm d}^2N}{{\rm d}\varphi'\,{\rm d}\!\cos\vartheta'}{\cal P}\bigr)\big|_{\varphi'=\delta_0+\frac{\pi}{2}\delta_{i2},\vartheta'=\sigma_i}=1$, where $\delta_{ij}$ is the Kronecker delta; cf. Fig.~\ref{fig:d2Nperpbyd2NbyP} for an illustration.
Employing the elliptical symmetry for $\vec{x}_0=0$ we then adopt the ansatz $\sigma(\varphi')=\frac{\sigma_1\sigma_2}{\sqrt{\sigma_2^2\cos^2(\varphi'-\delta_0)+\sigma_1^2\sin^2(\varphi'-\delta_0)}}$
to generalize this condition to arbitrary values of $\varphi'$, resulting in 
$\frac{{\rm d}^2N_\perp}{{\rm d}\varphi'\,{\rm d}\!\cos\vartheta'}/\bigl(\frac{{\rm d}^2N}{{\rm d}\varphi'\,{\rm d}\!\cos\vartheta'}{\cal P}\bigr)\big|_{\vartheta'=\sigma(\varphi')}=1$;
we have explicitly checked that for the considered cases this condition is indeed fulfilled to excellent accuracy.
Signal photons emitted in directions with $\vartheta'\geq\sigma(\varphi')$ can be detected with state-of-the-art technology;
we denote the total number of perpendicularly polarized photons fulfilling this requirement by $N_{\perp>\sigma}$.
In the last column of Tab.~\ref{tab:data} we give the number of perpendicularly polarized signal photons attainable per hour, assuming the probe pulse to comprise $N=10^{12}$ photons -- which matches the parameters of the European XFEL \cite{XFEL} -- and a repetition rate of $1$Hz. The repetition rate is limited by the high-intensity laser system: State-of-the-art $1$PW laser systems such as the Berkeley Lab Laser Accelerator (BELLA) \cite{BELLA} fire with a repetition rate of $1$Hz.

\begin{figure}
\center
\includegraphics[width=0.85\textwidth]{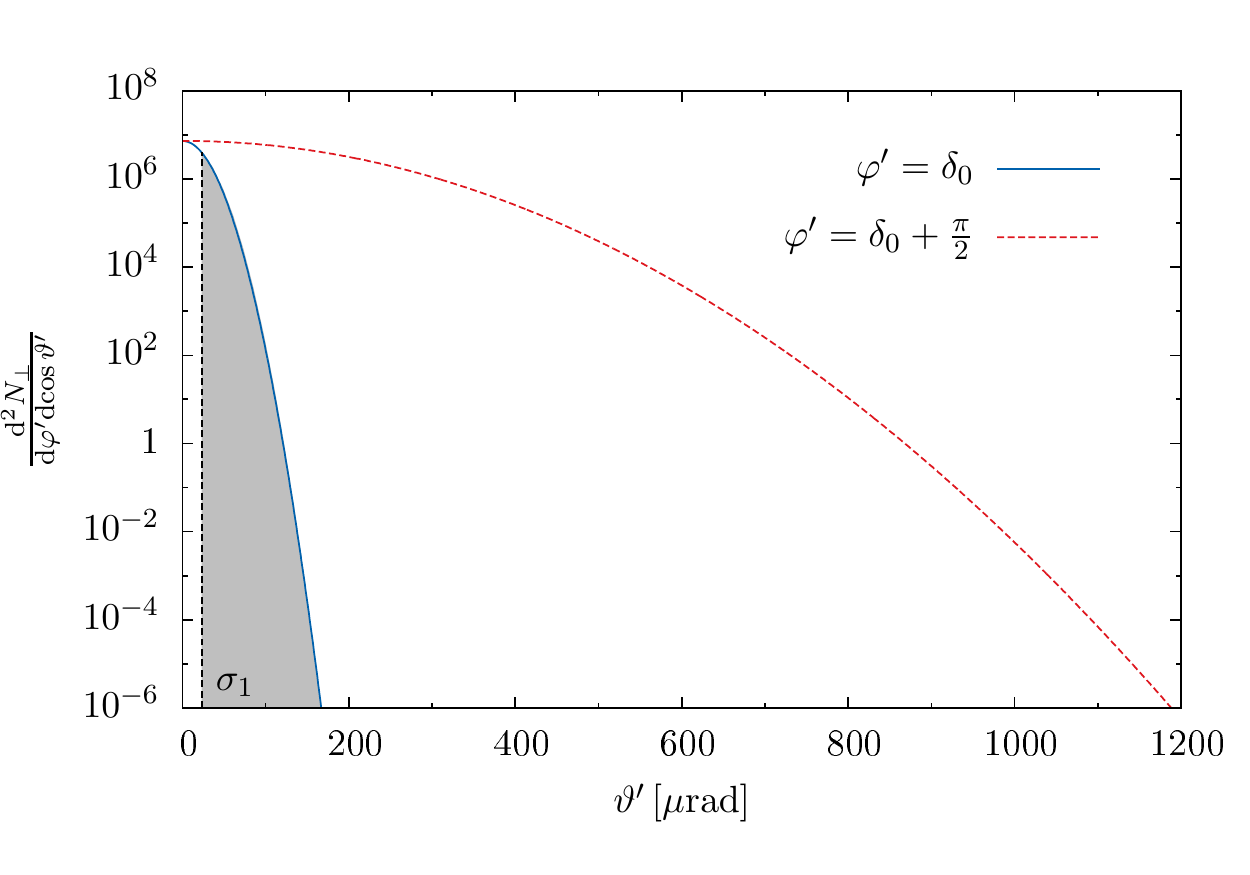}
\caption{Differential number of perpendicularly polarized signal photons $\frac{{\rm d}^2N_\perp}{{\rm d}\varphi'\,{\rm d}\!\cos\vartheta'}$ plotted as a function of $\vartheta'$ for a probe of asymmetric cross-section; waist parameters $\frac{w_1}{w_0}=3$ for $\varphi'=\delta_0$, and $\frac{w_2}{w_0}=\frac{1}{10}$ for $\varphi'=\delta_0+\frac{\pi}{2}$.
The segment of the blue (solid) curve highlighted in gray fulfills $\vartheta'\geq\sigma_1$, i.e., corresponds to signal photons emitted at an azimuthal angle of $\varphi'=\delta_0$ which could be detected employing state-of-the-art polarization purity measurements. An analogous regime also exits for the red (dashed) curve. However, it is only of academic relevance as it lies  far outside the depicted angle interval beyond $\vartheta'=\sigma_2=5129.38\mu$rad (cf. Tab.~\ref{tab:data}), where $\frac{{\rm d}^2N_\perp}{{\rm d}\varphi'\,{\rm d}\!\cos\vartheta'}$ has essentially dropped to zero. Correspondingly, no signal photons are to be detected in $\varphi'=\delta_0+\frac{\pi}{2}$ direction.
The two curves depicted here are continuously related by means of the angle $\varphi'$ (cf. main text).}
\label{fig:d2Nperpdphisdcosthetas}
\end{figure}

As obvious from Tab.~\ref{tab:data}, for small probe beam waists $w_i\lesssim w_0$ this criterion results in comparably large values for the corresponding $\sigma_i$.
Essentially no signal photons are emitted for such large values of $\sigma_i$, rendering the above criterion purely academic for the associated direction.
Contrarily, for larger probe beam waists $w_i\gtrsim w_0$ a substantial fraction of signal photons can be emitted outside $\sigma_i$, facilitating their detection with state-of-the-art polarimetry (see Fig.~\ref{fig:d2Nperpdphisdcosthetas}).
In Tab.~\ref{tab:data}, the ratio $N_{\perp>\sigma}/N_\perp$ is maximal for the largest probe beam cross-section considered.
Nevertheless, the total number of signal photons to be detected experimentally, $N_{\perp>\sigma}$, is largest for the asymmetric cross-section with $w_1=3w_0$ and $w_2=\frac{1}{10}w_0$. For this configuration the probe photons propagate close to the beam axis of the pump -- increasing the experienced pump field strength, triggering the effect -- in one direction and are spread out to sense its full transverse profile -- increasing the signals' angular spread, facilitating its detection with state-of-the-art-technology -- in the perpendicular direction.
Hence, this observation can be explained as resulting from a beneficial combination of these two effects.

Finally, note that a direct comparison of our results with the estimates for the number of perpendicularly polarized signal photons obtained in \cite{Karbstein:2015xra} shows that our new -- more realistic -- predictions tend to be somewhat smaller.
More specifically, adopting the same parameters for the case of $\frac{w_1}{w_0}=\frac{w_2}{w_0}=3$, \cite{Karbstein:2015xra} obtained $\frac{N_\perp}{N}=4.31\cdot10^{-14}$ (cf. last row of Tab.~(b) in \cite{Karbstein:2015xra}), to be compared with the value of $\frac{N_\perp}{N}=3.64\cdot10^{-14}$ determined here.
Let us emphasize again that the transverse probe profile was only indirectly accounted for in \cite{Karbstein:2015xra}: The result provided there relies on the assumption that $\{w_1,w_2\}\gg w_0$, which arguably might not be completely justified for the considered case, and partially explain the observed deviations. Another, but closely related source of discrepancies is that the transverse profile of the probe was assumed to be homogeneous in \cite{Karbstein:2015xra}, whereas it is of Gaussian type here. 

Reference~\cite{Karbstein:2015xra} also considered the case of an ideal line focus corresponding to $\frac{w_1}{w_0}=3$ and $\frac{w_1}{w_0}=0$, for which a value of $\frac{N_\perp}{N}=2.17\cdot10^{-13}$ was estimated (cf. last row of Tab.~(c) in \cite{Karbstein:2015xra}). Assuming $N=10^{12}$ probe photons per pulse and a repetition rate of $1$Hz, \cite{Karbstein:2015xra} predicted $N_{\perp>\sigma}\approx265$ per hour.
This might be compared with the case of $\frac{w_1}{w_0}=3$ and $\frac{w_1}{w_0}=\frac{1}{10}$ studied here, yielding $\frac{N_\perp}{N}=1.49\cdot10^{-13}$ and resulting in $N_{\perp>\sigma}\approx146$ per hour\footnote{Note that in the present work we use a slightly more refined criterion to determine the value of $\sigma(\varphi')$. In \cite{Karbstein:2015xra} the value of $\sigma$ (called $\vartheta_{\rm min}$ in \cite{Karbstein:2015xra}) is determined by demanding that the ratio of the total number of perpendicularly polarized signal photons emitted outside $\sigma$ and the total number of probe photons propagating into directions outside $\sigma$ is larger than $\cal P$. In the present work, we implement this criterion on the level of the differential photon numbers (cf. Sec.~\ref{sec:predictions}), which ensures that this condition is met for each value of $\vartheta'\geq\sigma$ individually.}.

\section{Conclusions and Outlook} \label{sec:Con&Out}

In this article we have extended the recent study~\cite{Karbstein:2015xra} to account also for the details of the x-ray probe.
One of these features is the finite transverse extent of the probe beam governed by the two independent waists $w_1$ and $w_2$, enabling us to consider elliptically shaped probe beam cross-sections of arbitrary orientation.
This has in particular permitted us to substantiate previous results for the limiting cases of probe beams either significantly narrower or wider than the beam waist of the pump beam \cite{Karbstein:2015xra}.
In addition, similarly to \cite{Schlenvoigt:2016} our results account for a finite impact, or more generally arbitrary spatiotemporal displacements of the foci.
Note however that our present study allows us to go beyond this study, which exclusively focuses on the polarization-flip signal and by construction does not account for scattering effects. In fact, we expect the latter effects giving rise to signal photons scattered out of the cone of the probe photon beam as crucial means to enhance the signal-to-background ratio in a discovery experiment of QED vacuum birefringence with high-intensity lasers, employing state-of-the-art technology.
Also note that even though in the present study we mainly focused on a counter-propagation geometry of the pump and probe laser pulses, our general results are valid for arbitrary collision geometries.

We are confident that our study will pave the way for a realistic study of vacuum birefringence, facilitating stringent quantitative predictions and optimizations of the signal in an actual experiment. Based on the new insights obtained here, a detailed study of a precise experimental scenario similar to \cite{Schlenvoigt:2016} has now become feasible. 

\acknowledgments

We are particularly grateful to Maria~Reuter for creating Fig.~\ref{fig:3d}
and are indebted to Holger~Gies for many stimulating discussions and useful comments on this manuscript.
F.K. would like to thank Matt~Zepf for many enlightening discussions and helpful explanations.
Moreover, inspiring discussions with Hendrik Bernhardt, Tom Cowan, Benjamin Grabiger, Malte~C.~Kaluza, Tino K\"ampfer, Robert L\"otzsch, Gerhard G. Paulus, Ingo Uschmann, Roland Sauerbrey, Hans-Peter Schlenvoigt, Kai-Sven Schulze, and cryogenic delights provided by Ingo~Uschmann are gratefully acknowledged.
Support by the DFG under grant No.~SFB-TR18 is gratefully acknowledged.

\appendix

\section{An approximate result for the number of signal photons}\label{app:Appendix}

When replacing $w(\rz)\to w_{\rm eff}={\rm const}.$ in the exponential of \Eqref{eq:E(x)^2approx}, which amounts to approximating the pump radius $w(\rz)$ as constant, 
and thereby neglecting any beam-widening effects as a function of $\rz$, also the $\rz$ integration in \Eqref{eq:M} can be performed analytically:
On the level of \Eqref{eq:M} this approximation amounts to replacing all the implicit dependences on $w(\rz)$ encoded in the functions \eqref{eq:j}-\eqref{eq:f} by the effective beam radius $w_{\rm eff}$. 
Correspondingly, the $\rz$ dependence of the integrand in \Eqref{eq:M} is only in terms of the overall factor of $(\frac{w_0}{w({\rm z})})^2$, as well as quadratic and linear terms in the exponential.
To perform the integration over $\rz$ we can then resort to the following identity for $A\geq0$,
\begin{equation}
 \int{\rm dz}\,\Bigl(\frac{w_0}{w({\rm z})}\Bigr)^2 {\rm e}^{-A\rz^2+(B+\ri C)\rz} = \frac{\pi}{2}\rz_R\,{\rm e}^{A\rz_R^2}\sum_{\ell=\pm1}{\rm e}^{\ell\rz_R(C-\ri B)}
 \biggl[1-{\rm erf}\Bigl(\rz_R\sqrt{A}+\tfrac{\ell}{2}\tfrac{C-\ri B}{\sqrt{A}}\,\Bigr)\biggr] , \label{eq:nowideningint}
\end{equation}
where ${\rm erf}(.)$ is the error function.
As the result is rather lengthy and does not allow for new insights we refrain from quoting its explicit expression here.
Even though this approximation does not account for beam widening effects, it correctly accounts for a finite focusing length along $\rz$, going along with a drop of the pump intensity $\sim(\frac{w_0}{w({\rm z})})^2=[1+(\frac{\rz}{\rz_R})^2]^{-1}$ with increasing distance from the focus.

The length scale $w_{\rm eff}$  can be interpreted as the effective beam radius of the pump in the interaction volume. It can be tuned such that the number of signal photons obtained with this approximation matches the full calculation.
Because of $w(\rz)\geq w_0$, the naive identification $w_{\rm eff}=w_0$ reduces the strong-field volume and thus generically leads to an underestimation of the number of signal photons:
Adopting $w_{\rm eff}=w_0$, the field strength squared of the pump scales as $\sim{\rm exp}\{{-2\frac{\rx^2+\ry^2}{w_0^2}}\}$ and falls off faster than the full result accounting for the beam widening which scales as $\sim{\rm exp}\{{-2\frac{\rx^2+\ry^2}{w^2(\rz)}}\}$.
For these reasons we expect to find $w_{\rm eff}\geq w_0$.

Subsequently, we stick to the case of counter-propagating pump and probe laser pulses as discussed in Sec.~\ref{subsec:counterprop}.
Here, we fix $w_{\rm eff}$ by demanding that for given parameters of the pump and probe laser pulses we have $\frac{{\rm d}N_\perp^{\rm approx}}{{\rm d}\cos\vartheta'}\big|_{\vartheta'=0}=\frac{{\rm d}N_\perp}{{\rm d}\cos\vartheta'}\big|_{\vartheta'=0}$, where $\frac{{\rm d}N_\perp}{{\rm d}\cos\vartheta'}\equiv\int_0^{2\pi}{\rm d}\varphi'\frac{{\rm d}^2N_\perp}{{\rm d}\varphi'{\rm d}\cos\vartheta'}$.
Note that the calculation for $\vartheta'=0$ is considerably simpler than for $\vartheta'\neq0$, as the dependence on the azimuthal angle $\varphi'$ drops out in this limit.
In turn the integration over $\varphi'$ amounts to a simple multiplication with a factor of $2\pi$.
The approximative expression for the differential number of perpendicularly polarized signal photons attainable from Eqs.~\eqref{eq:Mcp} and \eqref{eq:nowideningint} is given by
\begin{multline}
{\rm d}^3N_{\perp}^{\rm approx}
 \approx
m^4\frac{{\rm d}^3k'}{(2\pi)^3}\,\rk'(w_{\rm eff}^2\rz_R\tau)^2(1+\cos\vartheta')^2\,
\alpha\,\Bigl(\frac{\pi}{120}\Bigr)^2\Bigl(\frac{e{\frak E}_0}{2m^2}\Bigr)^2\Bigl(\frac{e{\cal E}_0}{2m^2}\Bigr)^4\, \\
 \times\frac{(w_1w_2)^2}{w_{\rm eff}^4+2w_{\rm eff}^2(w_1^2+w_2^2)+4(w_1w_2)^2}\,\frac{1}{1+\frac{1}{2}(\frac{\tau}{T})^2}\, \\
 \times {\rm e}^{-\frac{1}{2}(w_{\rm eff}\rk'\sin\vartheta')^2\frac{w^2_{\rm eff}{\rm w}^2(\varphi')+2(w_1w_2)^2}{w_{\rm eff}^4+2w_{\rm eff}^2(w_1^2+w_2^2)+4(w_1w_2)^2}}
 {\rm e}^{-4\frac{(w_{\rm eff}^2+2w_2^2)(\hat{\vec{a}}\cdot\vec{x}_0)^2
+(w_{\rm eff}^2+2w_1^2)(\hat{\vec{b}}\cdot\vec{x}_0)^2}{w_{\rm eff}^4+2w_{\rm eff}^2(w_1^2+w_2^2)+4(w_1w_2)^2}+\frac{8}{T^2}\frac{(2\rz_R)^2-(\rz_0+t_0)^2}{1+\frac{1}{2}(\frac{\tau}{T})^2}} \\
\times\biggl|\sum_{q=\pm1} {\rm e}^{-\frac{2\tau^2}{1+\frac{1}{2}(\frac{\tau}{T})^2}(\frac{q\omega-\rk'}{8})^2}
\sum_{\ell=\pm1} {\rm e}^{\ell\rz_R\bigl[\rk'(1-\cos\vartheta')+\frac{2(q\omega-\rk')}{1+\frac{1}{2}(\frac{\tau}{T})^2}\bigr]} 
 {\rm e}^{-{\rm i}\bigl[
 \frac{(q\omega-\rk')+\ell\rz_R(\frac{4}{T})^2}{1+\frac{1}{2}(\frac{\tau}{T})^2}(\rz_0+t_0)+q\psi_0\bigr]} \\
 \times\biggl[1-\ell\,{\rm erf}\biggl(\tfrac{T}{4}\tfrac{(q\omega-\rk')+\ell\rz_R(\frac{4}{T})^2+\frac{\rk'}{2}(1-\cos\vartheta')[1+\frac{1}{2}(\frac{\tau}{T})^2]}{\sqrt{1+\frac{1}{2}(\frac{\tau}{T})^2}}
 -\ri \tfrac{2\frac{\rz_0+t_0}{T}}{\sqrt{1+\frac{1}{2}(\frac{\tau}{T})^2}}\biggr)\biggr]\biggr|^2 \,. \label{eq:d3Nperpapprox}
\end{multline}

Exemplarily sticking to the case of $\frac{w_1}{w_0}=3$ and $\frac{w_2}{w_0}=\frac{1}{10}$, discussed extensively in Sec.~\ref{sec:predictions}, we infer $\frac{w_{\rm eff}}{w_0}\approx1.058$.
To allow for an easy comparison, we contrast our approximate results obtained from \Eqref{eq:d3Nperpapprox} to the corresponding exact results derived in Sec.~\ref{sec:predictions} (cf. in particular Tab.~\ref{tab:data}). We find
\begin{table}[h]
\begin{tabular}{|c||c|c|c|c|c|c|}
 \hline
   & $\frac{N_\perp}{N}$  & $\frac{N_{\perp>\theta}}{N_\perp}$ & $\sigma_1[\mu{\rm rad}]$ & $\sigma_2[\mu{\rm rad}]$ & $\frac{N_{\perp>\sigma}}{N_\perp}\phantom{\Big.^{|}}$\hspace*{-1mm} & $\frac{N_{\perp>\sigma}}{\rm h}$ \\
 \hline
 \hline
  Exact Result & $1.49\cdot10^{-13}$ & $68.5\%$ & $23.14$ & $5129.38$ & $27.2\%$ & $145.74$ \\
  Approximation &  $1.49\cdot10^{-13}$ & $68.5\%$ & $23.13$ & $5331.03$ & $27.2\%$ & $146.64$ \\
 \hline
 \end{tabular}\ .
\end{table}

\noindent Here, we encounter tiny deviations in the results for $\sigma_1$ and somewhat larger ones for $\sigma_2$, which is anyway outside the regime where significant signal photon contributions are to be expected.
To the precision shown, the results of the approximation are in good agreement with the exact results for the signal photon numbers.

For completeness, note that this approximation does not account for all the effects contained in the full result, Eqs.~\eqref{eq:d3Nperpcp} and \eqref{eq:Mcp}. In particular, it
does not resolve azimuthal asymmetries in the differential number of signal photons expected to be encountered for $w_1=w_2$ but $\hat{\vec{a}}\cdot\vec{x}_0\neq0$ or $\hat{\vec{b}}\cdot\vec{x}_0\neq0$:
For $w(\rz)\to w_{\rm eff}$, the terms inducing such effects in \Eqref{eq:Mcp} become a pure phase factor, and thus drop out upon taking the modulus of ${\cal M}_0^{\rm approx}$.

\end{document}